\documentclass[aps,prb,twocolumn,amsmath,amssymb,superscriptaddress]{revtex4}
\usepackage{graphicx}
\usepackage{dcolumn}
\usepackage{hyperref}
\usepackage[usenames,dvips]{color}
\usepackage[usenames,dvipsnames]{xcolor}
\usepackage{bm}
\begin{document}

\providecommand{\com}[1]{{\color{green}[\emph{#1}]}}
\newcommand{\corout}[1]{{\color{red}\sout{#1}}}
\newcommand{\cor}[1]{{\color{red} #1}}
\newcommand{\cob}[1]{{\color{blue} #1}}
\newcommand{\highlight}[1]{{\color{Violet} #1}}
\newcommand{\someref}{{\bf\color{Orange}[ ? ]}}

\newcommand{\corSt}[1]{{\color{Orange} #1}}

\title {Inelastic Decay of Electrons in the Shockley-type Metal-Organic Interface States}

\author{S.~S.~Tsirkin}
 \affiliation{Donostia International Physics Center (DIPC), 20018 San Sebasti\'an/Donostia, Basque Country, Spain\\}
 \affiliation{Tomsk State University, 634050, Tomsk, Russia\\}
 \affiliation{Saint Petersburg State University, Saint Petersburg, 198504, Russia\\}

\author{N.~L.~Zaitsev}
 \affiliation{Philipps-Universit\"{a}t Marburg, D-35032, Marburg, Germany}

\author{I.~A.~Nechaev}
 \affiliation{Donostia International Physics Center (DIPC), 20018 San Sebasti\'an/Donostia, Basque Country, Spain\\}
 \affiliation{Tomsk State University, 634050, Tomsk, Russia\\}

\author{R. Tonner}
\affiliation{Philipps-Universit\"{a}t Marburg, D-35032, Marburg, Germany}

\author{U. H\"{o}fer}
\affiliation{Philipps-Universit\"{a}t Marburg, D-35032, Marburg, Germany}

\author{E.~V.~Chulkov}
 \affiliation{Donostia International Physics Center (DIPC), 20018 San Sebasti\'an/Donostia, Basque Country, Spain\\}
 \affiliation{Tomsk State University, 634050, Tomsk, Russia\\}
 \affiliation{Departamento de F\'{\i}sica de Materiales UPV/EHU, Facultad de Ciencias Qu\'{\i}micas, UPV/EHU, Apdo. 1072, 20080 San Sebasti\'an/Donostia, Basque Country, Spain\\}
 \affiliation{Centro de F\'{\i}sica de Materiales CFM - MPC, Centro Mixto CSIC-UPV/EHU, 20080 San Sebasti\'an/Donostia, Basque Country, Spain\\}
 \affiliation{Saint Petersburg State University, Saint Petersburg, 198504, Russia\\}

\date{\today}

\begin{abstract}
We present a theoretical study of  lifetimes of interface states (IS) on metal-organic interfaces  PTCDA/Ag(111),
NTCDA/Ag(111), PFP/Ag(111), and PTCDA/Ag(100), describing and explaining the recent experimental data. 
By means of unfolding the band structure of one of the interfaces under study onto the Ag(111) Brillouin zone we demonstrate, that the 
Brillouin zone folding upon organic monolayer deposition plays a minor role in the phase space for electron decay, and hence weakly affects the resulting lifetimes. 
The presence of the unoccupied molecular states below the IS gives a small contribution to the IS decay rate mostly determined by the change of the phase space of bulk states upon the energy shift of the IS. The calculated lifetimes follow the experimentally observed trends.  In particular, we explain the trend of the unusual increase of the IS lifetimes  with rising temperature.
\end{abstract}

\pacs{
73.20.-r,   
79.60.Dp     
78.47.-p,   
}

\maketitle

\section{Introduction}
\label{sec:Intro}

Many well-defined interfaces between organic semiconductors and metals exhibit interface-specific
electronic states that exist independently of the detailed molecule-substrate interaction.
\cite{Szyman05pss,Marks14jelsp,Ilyas14prb,Zhao14acsnano}
 Like the surface states (SSs) of clean metals they are a consequence of the breaking of translational
symmetry perpendicular to the interface.
 In the Shockley-type interface states (ISs) that have been identified at Ag(111) and Ag(100) interfaces
the electrons are able to move almost freely parallel to the interface whereas the local charge
density in the vicinity of the first molecular layer is strongly corrugated and resembles that of
molecular orbitals.
 These general properties of the states have been revealed in recent experimental and theoretical
works.\cite{Temirov06nat, Schwalb08prl, Yang08jpcc,  Scheybal09prb, Dyer_NJP_2010, Zaitsev10,
Sachs09jcp, Schwalb10epj, Marks_PRB_2011, Zaitsev_PRB_2012, Galbraith_JPCL_2014, Caplins14jpcl}
 For a few systems also the important factors that determine their electronic structure could be
investigated.\cite{Marks_PRB_2011,Caplins14jpcl}

 The dynamics of electronic decay and electron transfer processes at interfaces that involve these
states, however, is not well understood.
 Previous two-photon photoemission (2PPE) experiments have measured lifetimes between 10 and 200~fs for electrons excited
into normally unoccupied interface states above the Fermi level.\cite{Schwalb08prl, Marks_PRB_2011,
Galbraith_JPCL_2014,Caplins14jpcl}
 From these short lifetimes a large overlap of the wave function with the metal has been concluded.
These conclusions, although confirmed by density functional calculations,\cite{Dyer_NJP_2010,
Zaitsev10, Marks_PRB_2011, Zaitsev_PRB_2012, Galbraith_JPCL_2014} are based on very simplistic assumptions on the
nature of electronic decay processes at such an interface.
 Many-body calculations, such that exist for surfaces states of clean metal surfaces,\cite{Echenique04ssr} have not been performed so far.

 In this publication, we make a first step in this direction. 
We perform a theoretical study of the decay of electrons in the interface states  formed at 
the interfaces of silver with ordered monolayers of 
such organic molecules as perylene-3,4,9,10-tetracarboxylic acid dianhydride (PTCDA), 
naphthalene-1,4,5,8-tetracarboxylic acid dianhydride (NTCDA) and perfluoropentacene (PFP).
First, on an equal footing we perform ab initio density functional calculations of the electronic structure of all the studied interfaces,
of which the PFP/Ag(111) interface is published for the first time here.
 Our calculations show that the hybridization of molecular and metallic
states is small in the region of the projected band gap of the metal. New elastic decay channels,
which in principle could open up due to reduced translational symmetry of the organic overlayers,
will thus only have a weak influence on the electron decay of the interface state. This allows us
to focus on the inelastic decay of electrons excited to the IS.
 We calculate the corresponding lifetimes in the self-energy formalism of many-body theory using
the GW approximation\cite{hedin65}.
 In order to make the calculations feasible, we use one-dimensional model potentials for an
approximate description of the electronic structure of the interfaces. 
 The potentials are based on the so-called Chulkov potential\cite{Chulkov_SS_1997,Chulkov_SS_1999} of clean surfaces. They are modified
in order to reproduce the experimentally observed energy upshift of the Shockley surface state in
the presence of the organic overlayers. Comparison with the ab initio calculations allows us to
judge how well the potentials reproduce the probability density of the states perpendicular to the
interfaces.
 The results are in good overall agreement with the experimental data for PTCDA	/Ag(111),
NTCDA/Ag(111), PFP/Ag(111), and PTCDA/Ag(100). Calculated lifetimes are generally longer than
experimental ones but agree well with experimental trends.

\section{Ab initio calculation of electronic structure of interfaces}

First, in order to have a detailed information about the electronic structure of the interfaces under study, we performed \textit{ab-initio} calculations within the periodic slab geometry. We used the OPENMX (version 3.7) code\cite{_openmx_37}, which is based on density functional theory and the linear combination of localized pseudoatomic orbital (LCPAO) method.\cite{ozaki_variationally_2003,ozaki_numerical_2004,ozaki_efficient_2005} We applied the generalized gradient approximation (GGA) of Ref.~\onlinecite{perdew_generalized_1996} for the exchange-correlation functional. Also we exploited norm-conserving pseudopotentials\cite{troullier_efficient_1991} in order to replace deep core potentials by shallow ones. For silver atoms, we set basis functions to Ag7.0$-s2p2d2f1$, while for hydrogen, carbon, and oxygen atoms we use  H6.0$-s2p1$, C6.0$-s2p2d1$ and O6.0$-s2p2d1$, respectively. On the example of silver, this notation means that two primitive orbitals for each \emph{s}, \emph{p}, and \emph{d} orbital and one primitive orbital for the \emph{f} orbital were used for representation of the basis functions with the cutoff radius of 7.0 Bohr.

\begin{figure}[t]
\begin{center}
\includegraphics[angle=0, width=\columnwidth]{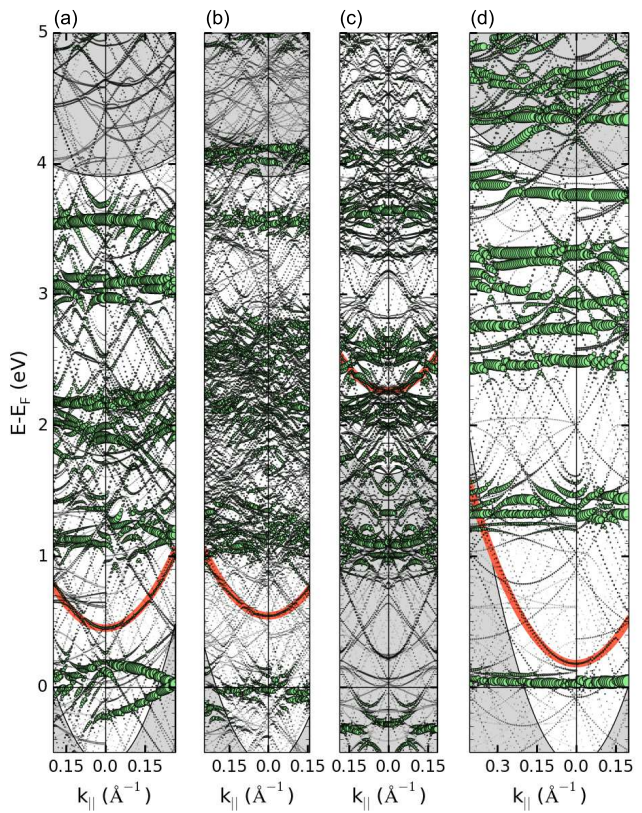}
\caption{Band structure of (a) NTCDA@Ag(111) in the relaxed phase, (b) PTCDA@Ag(111), (c) PTCDA@Ag(100), and (d) PFP@Ag(111). The width of the fat bands reflects the extent of their localization in the vicinity of the molecular ML. The red curves highlight the interface state originated from the Shockley surface state (resonance) of the respective bare silver surface.}
 \label{fig:LCPAO}
\end{center}
\end{figure}

To improve the description of the surface state (SS), we used the enlarged cutoff radius (9.0 Bohr) for silver atoms in the uppermost layers of the slab. The latter contains 10 silver layers together with the molecular monolayer (ML) attached on one side of the silver film. The vertical distances between the Ag(111) surface and the carbon atoms of  NTCDA,\cite{stadler_molecular_2007} PTCDA,\cite{hauschild_normal-incidence_2010} and  PFP\cite{duhm_influence_2010} MLs are taken to be equal to its experimental values. The oxygen atoms of NTCDA and PTCDA monolayers on Ag(111) are fixed at the same distance as carbon ones. The optimized geometry of PTCDA/Ag(100) found in Ref.~\onlinecite{Galbraith_JPCL_2014} is used in our calculation. The real-space grid for numerical integration and solution of the Poisson equation was specified by the energy cutoff of 250 Ry. The total-energy convergence was better than 0.027 meV. The surface Brillouin zone (SBZ) of the supercell was sampled with a $3\times3\times1$ mesh of \textbf{k}-points.

As seen in Fig.~\ref{fig:LCPAO}, due to the unit cell of the interfaces, which is larger than the ($1\times1$) cell naturally used for the bare surfaces, the SBZ becomes smaller, and, consequently,
the metal bands of the initial SBZ corresponding to the ($1\times1$) unit cell get folded into the reduced SBZ, hybridizing at that with the orbitals of the molecular ML. This leads to a surface band structure that does not exhibit the projected band gap at the $\bar{\Gamma}$ point any more. To restore the ($1\times1$) unit-cell representation of the interface electronic structure, we performed an unfolding procedure using the BandUP code,\cite{unfolding1,unfolding2,unfolding3} based on the method by Popescu and Zunger. \cite{Popescu}
Upon the unfolding procedure each electronic state $m\mathbf{K}$ (where $m$ is the band index and $\mathbf{K}$ is the wavevector in the SBZ of the interface) is projected onto a set of corresponding points $\mathbf{k}_i$ in the unfolded Brillouin zone (UBZ) of the ($1\times1$) unit cell, resulting the weights $W_{m\mathbf{K}}(\mathbf{k}_i)$.
The BandUP code deals with wavefunctions, expressed in the plane-wave basis set. 
Thus we perform calculations with the  VASP code,\cite{VASP1,VASP2} based on the plane-wave basis, employing the projector-augmented wave (PAW) method. \cite{PAW1,PAW2}
The exchange-correlation was calculated within the  GGA,\cite{perdew_generalized_1996} like in the LCPAO method. The energy cut-off was fixed at 350 eV and a 6$\times$4$\times$1 Monkhorst-Pack 
grid of k-points was used.
These calculations are notably more time-consuming than the LCPAO-calculations, thus we restrict ourselves to the case of NTCDA@Ag(111).

The band-structure calculations indicate that  the deposition of a ML on top of the silver surface influences the partly occupied SS in a qualitatively similar manner for different interfaces. In the cases of Ag(111) surface the SS is shifted to a higher energy and transformed into the interface state (IS). In the case of the Ag(100) surface, the unoccupied Shockley resonance (SR) is also up-shifted in energy, which changes its character to a distinct electronic state. The magnitude of the upshift depends on the molecular type, coverage, surface orientation, and adsorption geometry\cite{Schwalb08prl, Zaitsev10, Dyer_NJP_2010, Marks_PRB_2011, Zaitsev_PRB_2012, Galbraith_JPCL_2014} (see Table~\ref{tab:IS_energies}).  Additionally, the bare-surface electronic structure is filled up by the molecular-derived weakly dispersive states, which are thought to have an effect on electron dynamics in higher lying states.

In the unfolded bandstructure (Fig \ref{fig:unfolded}a) one can see a clearer picture how  the deposition of an NTCDA ML modifies the electronic structure of the Ag(111) surface. 
Below $-3$ eV one can see a set of $d$-bands, which are not affected by the molecular ML. At higher energies the bulk-derived $s-p$ bands forming the gap at the $\overline{\Gamma}$ 
point at $-1.2\textrm{ eV}<E<3.6\textrm{ eV}$, also have practically the same energies for surfaces with and without NTCDA. The main effect of NTCDA ML consists in the transformation of 
the bonding and anti-bonding surface states (at -0.17 and +0.08 eV) of the bare 10-layer Ag(111) slab into the SS of the clean side of the slab (at -0.06 eV) and the IS  of the slab with NTCDA at $E_{IS}=0.5$ eV. The latter value agrees with the result of LCPAO calculations $E_{IS}=0.4$ eV. The molecular states of NTCDA are unfolded to different points of the UBZ with small weights, thus forming a weak background.
Note, that the spectral weight of the molecular-derived states is quite small, hence one could expect them to produce a rather minor contribution to decay processes. This question will be addressed in more detail later in Sec. \ref{sec:MLpot}.

\begin{table*}
\caption{\label{tab:IS_energies} Experimental and theoretical values of the IS energy $E_{IS}$ (in eV) and lifetimes $\tau_{IS}$ (in fs).
Theoretical values of $\tau_{IS}$ are given for the $E$-shifted / $V$-shifted scheme, accounting for inelastic electron-electron scattering only.} 
\begin{ruledtabular}
\begin{tabular}{cccccc}
  &             & PTCDA@Ag(111) &  PTCDA@Ag(100)   & NTCDA@Ag(111)   & PFP@Ag(111) \\
  \hline
$E_{IS}$  &Experiment   & $0.57\pm0.02$\footnotemark[1] &  $2.25\pm0.03$\footnotemark[2]   &  $0.38\pm0.02$\footnotemark[1]  & 0.1-0.2\footnotemark[3]  \\
  &Theory       & 0.55          &  2.26            &  0.40           & 0.17  \\
  \hline
$\tau_{IS}$  &Experiment   & $53\pm3$\footnotemark[1] &  $3\le\tau\le18$\footnotemark[2]   &  $115\pm10$\footnotemark[1]  &  --- \\
  &Theory       & 110 / 270          &  --- /  24           &  250 / 500          &   1280 / 1850  
\end{tabular}

\footnotetext[1]{From Ref.~\onlinecite{Marks_PRB_2011}}
\footnotetext[2]{From Ref.~\onlinecite{Galbraith_JPCL_2014}}
\footnotetext[3]{From Ref.~\onlinecite{Galbraith_PHD}}

\end{ruledtabular}
\end{table*}

\section{GW approximation \label{sec:theor:meth}}

The lifetimes of ISs are calculated within the GW formalism.\cite{hedin65} 
As far as a fully ab initio calculation of lifetimes for the interface under consideration is a big challenge so far, one needs a model for the description of the electronic structure of these systems.
Such a model can be based on the following propositions. First, the origin of the IS experimentally observed in the aforementioned interfaces is attributed to the upshifted SS (SR) of the bare surfaces. (See e.g. Refs \onlinecite{Schwalb08prl, Dyer_NJP_2010, Zaitsev10, Marks_PRB_2011, Zaitsev_PRB_2012, Galbraith_JPCL_2014}) 
The properties of the resulting IS (the bulk penetration, near-surface localization, dispersion in $\mathbf{k}_{||}$, etc.) is similar to a Shockley-type state residing in the projected band gap of the bare (111) surface. 
Second, for IS  the decay phase space is thought to be mainly provided by the projected Ag bulk states and is not affected by the band folding caused by using the large interface unit cell instead of the (1$\times$1) unit cell of the bare Ag surfaces (see Fig. \ref{fig:unfolded}a).
Third, due to the rather small spectral weight of the molecule-derived states in the unfolded electronic surface structure, their contribution to the decay can be neglected. We reduce thus the problem to a study of electron lifetimes for a silver-like surface with a surface state modified in a way to reproduce the energy and dispersion of the considered IS.

The mentioned simplifications allow us to use a one-dimensional (1D) potential model\cite{Chulkov_SS_1997, Chulkov_SS_1999, Kliewer_Science_2000, Vitali_SSL_2003} in description of the decay of IS electrons. In such a model the one-electron pseudopotential $V(z)$ depends only on the $z$ coordinate, being constant in the $xy$ plane.
The $z$ axis is directed perpendicular to the surface outside the metal. The position of $z=0$ corresponds to the plane of the topmost Ag atomic layer on the side where the ML is attached. The forms of the potential $V(z)$ will be discussed further in sections \ref{sec:shiftedSS},    \ref{sec:MLpot}.

The method of calculation of lifetimes in the $GW$ approximation using a 1D pseudopotential is described in detail elsewhere,\cite{Echenique04ssr} and here we give just a brief overview, indispensable for understanding of the further discussion of the results. Within this formalism the decay rate $\Gamma_\textrm{e-e}$ of the electronic state in band $i$ with wave function
$\Psi_{\mathbf{k}i}(z,\mathbf{r}_{||})=\varphi_{i}(z)e^{i\mathbf{k}\mathbf{r}_{||}}$, energy $E_{\mathbf{k}i}=\epsilon_i+k^2/(2m_i^{\ast})$, and wavevector $\mathbf{k}$ is obtained as the projection of the imaginary part of the self-energy operator $\Sigma$ onto this state:
\begin{eqnarray}
\Gamma_\textrm{e-e}&=&-2\langle \Psi_{\mathbf{k}i} |
\mathrm{Im}\Sigma|\Psi_{\mathbf{k}i}\rangle\\
&=&-2\sum_{j}\int dz dz' M_{ij}(z,z')\nonumber\\
&\times&\int\frac{d\mathbf{q}}{(2\pi)^2}\left[1-f_{\mathbf{q}j}\right] \theta(E_{\mathbf{k}i}-E_{\mathbf{q}j}) \nonumber\\
&\times&\mathrm{Im}W(z,z';\mathbf{k}-\mathbf{q},E_{\mathbf{k}i}-E_{\mathbf{q}j}) \nonumber.
 \label{GammaEE}
\end{eqnarray}
Here $f_{\mathbf{q}j}$ is the Fermi factor and the wave-function product $M_{ij}(z,z') = \varphi_{i}(z)\varphi_{i}(z')\varphi_{j}(z) \varphi_{j}(z')$ is determined by the real eigenfunction $\varphi_{i}(z)$ being the solution of the one-dimensional Schr\"odinger equation
\begin{equation}
\label{schrodinger}
\left[-\frac{1}{2}\frac{{\rm d}^2}{{\rm d}z^2} + V(z)\right] \varphi_{i}(z) = \epsilon_i \varphi_{i}(z)
\end{equation}
with the respective eigenvalue $\epsilon_i$. In Eq.~(\ref{GammaEE}), the self-energy is represented by the first term of the expansion in terms of the screened Coulomb interaction $W$, which is calculated within the random phase approximation. Thus, the many-body decay rate is
determined by three main factors:
   (i) the phase space of the final states ($\mathbf{q}j$),
  (ii) the overlap between the wave functions of the initial and final
       states, and
 (iii) the magnitude of the imaginary part of the screened Coulomb
       interaction $\mathrm{Im}W$.
The latter is given in linear response theory by
\begin{eqnarray}\label{W_RPA}
W(z,z';\mathbf{q},\omega) &=& v_c(z,z';\mathbf{q}) + \int {\rm d} z_1 {\rm d} z_2 v_c(z,z_1;\mathbf{q})\nonumber\\ &\times&\chi(z_1,z_2;\mathbf{q},\omega)v_c(z_2,z';\mathbf{q}),
\end{eqnarray}
where $v_c(z,z';\mathbf{q})=2\pi e^{-q|z-z'|}/q$ is the 2D Fourier transform of the bare Coulomb interaction, and $\chi$ is the density-density response function of interacting electrons, which is evaluated from the  equation
\begin{eqnarray}
\chi(z,z';\mathbf{q},\omega) &=& \chi^0(z,z';\mathbf{q},\omega) + \int {\rm d} z_1 {\rm d} z_2 \chi^0(z,z_1;\mathbf{q}, \omega)\nonumber\\
&\times&v_c(z_1,z_2;\mathbf{q})\chi(z_2,z';\mathbf{q},\omega).
\label{Dyson_RPA}
\end{eqnarray}
Here $\chi^0(\mathbf{r}_1,\mathbf{r}_2;\omega)$ is the density-density response function of a non-interacting electron system:
\begin{equation}\label{x0_RPA}
\chi^0(z,z';\mathbf{q},\omega)=\sum_{ij}\int\frac{d\mathbf{k}}{(2\pi)^2}
\frac{(f_{\mathbf{k}j}-f_{\mathbf{k}+\mathbf{q}j})M_{ij}(z,z')}{\omega +
E_{\mathbf{k}j}-E_{\mathbf{k}+\mathbf{q}i}+i\eta}
\end{equation}
with $\eta$ being an infinitesimally small positive constant.

\section{Model of shifted surface state \label{sec:shiftedSS}}

As noted above, the one-dimensional (1D) potential model may be used for the description of the decay of IS electrons.
We start with the pseudopotential, introduced in Refs. \onlinecite{Chulkov_SS_1997} and \onlinecite{Chulkov_SS_1999} and a set of parameters that ensure the proper description of the surface electronic structure of the bare Ag surfaces. Further we need to modify the model to reproduce the energy of the IS, which is presented by the SS, shifted to higher energies, together with the energy of the $n=1$ IPS and the gap edges..

\begin{figure}[t]
\begin{center}
\includegraphics[angle=0, width=\columnwidth]{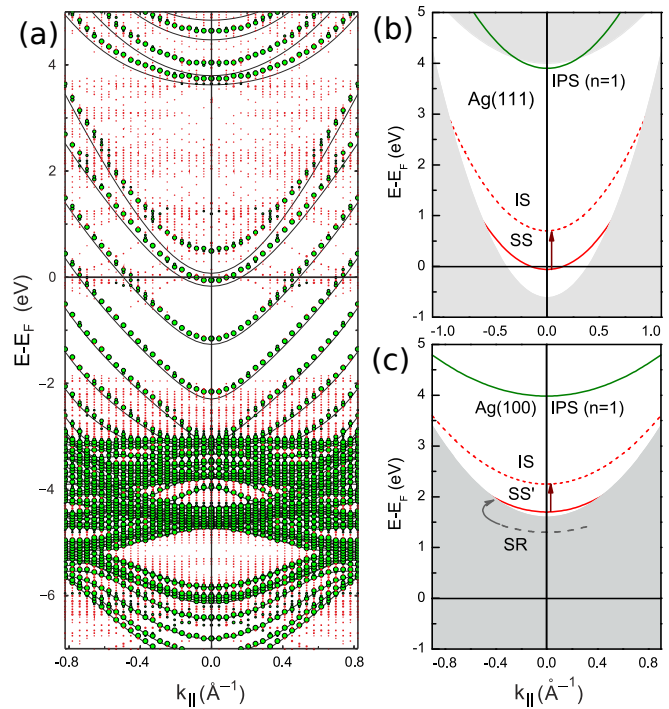}
\caption{
(a) Unfolded band structure of NTCDA@Ag(111) (dots) in comparison with the bare Ag(111) surface. The size of the symbol reflects the weight $W_{m\mathbf{K}}(\mathbf{k}_i)$ of the state in the unfolded BZ. The green (red) symbols denote the states with $W_{m\mathbf{K}}(\mathbf{k}_i)\ge0.1$ ($W_{m\mathbf{K}}(\mathbf{k}_i)<0.1$).
Sketch of the scheme realized in the model of the $V$-shifted surface state for Ag(111) (b) and Ag(100) (c). The SS is shifted towards higher energies by modifying the 1D pseudopotential of the bare Ag surfaces within the near-surface region. In the case of Ag(100), we start from the Shockley resonance (SR).
}
 \label{fig:unfolded}
\end{center}
\end{figure}

A way to achieve the shift of the SS, is to change the corresponding energies $\epsilon_i$, entering Eqs. (\ref{GammaEE}), (\ref{W_RPA}), (\ref{Dyson_RPA}) and (\ref{x0_RPA}) "by hand", while leaving the wavefunctions $\varphi_i(z)$ unchanged. 
 Hereafter, we refer to such a scheme as to the one of the {\it $E$-shifted surface state}. 
In this case we do not change the overlap between the IS and the bulk states, while the phase space and to a certain extent the screened interaction are modified.
Another way is to  tune the parameters of the 1D potential in the near-surface region in a way, which provides a shift of the surface-state energy towards higher values at the $\bar{\Gamma}$ point (see Fig.~\ref{fig:unfolded}b,c). 
This scheme will be referred to as the one of the {\it $V$-shifted surface state}.
Here, all the aforementioned factors that determine the inelastic decay are affected. 
In both schemes the energy of the IPS ($n=1$) and the surface barrier are held unchanged.

\begin{figure}[tbp]
\begin{center}
\includegraphics[angle=0, width=\columnwidth]{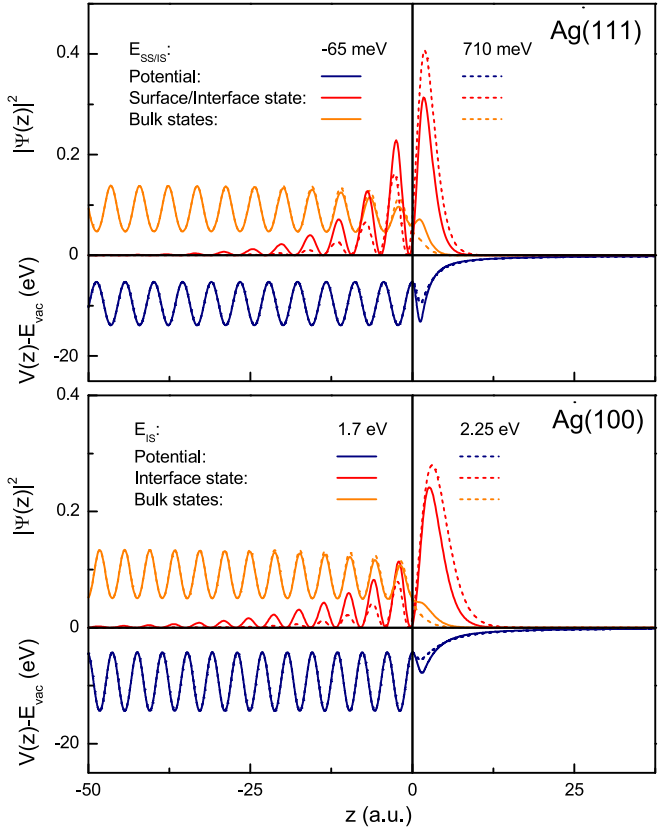}
\caption{One-dimensional potential and respective wave functions for the surface (interface) states and  bulk states in the case of the Ag(111) and Ag(100) surfaces at different energies of the shifted surface state ($E_{SS/IS}$). The bulk charge density was normalised to fit the same scale as SS/IS. }
 \label{fig:pseudo}
\end{center}
\end{figure}

\begin{figure}[tbp]
\begin{center}
\includegraphics[angle=0, width=\columnwidth]{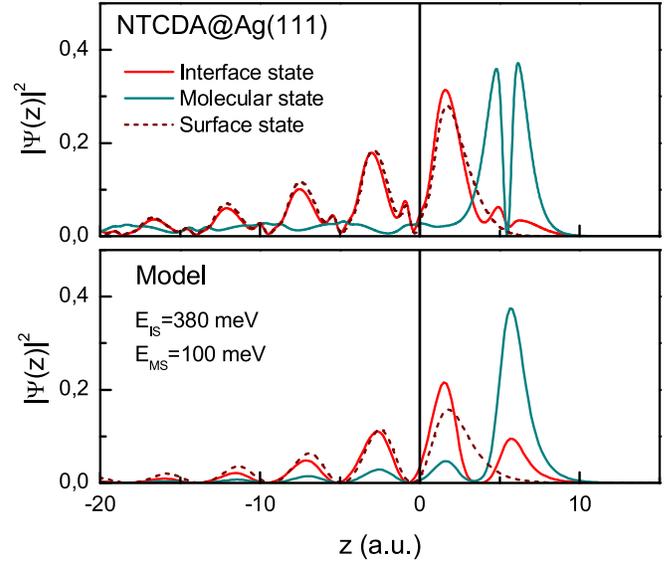}
\caption{Top: \textit{Ab-initio} calculated wave functions of the interface and former LUMO states as compared with that of the surface state of the bare Ag(111) surface in the case of NTCDA@Ag(111). Bottom: The same wave functions as obtained within the ML-pseudopotential model (see the text).}
 \label{fig:MLpot}
\end{center}
\end{figure}

\begin{figure}[tbp]
\begin{center}
\includegraphics[angle=0, width=0.8\columnwidth]{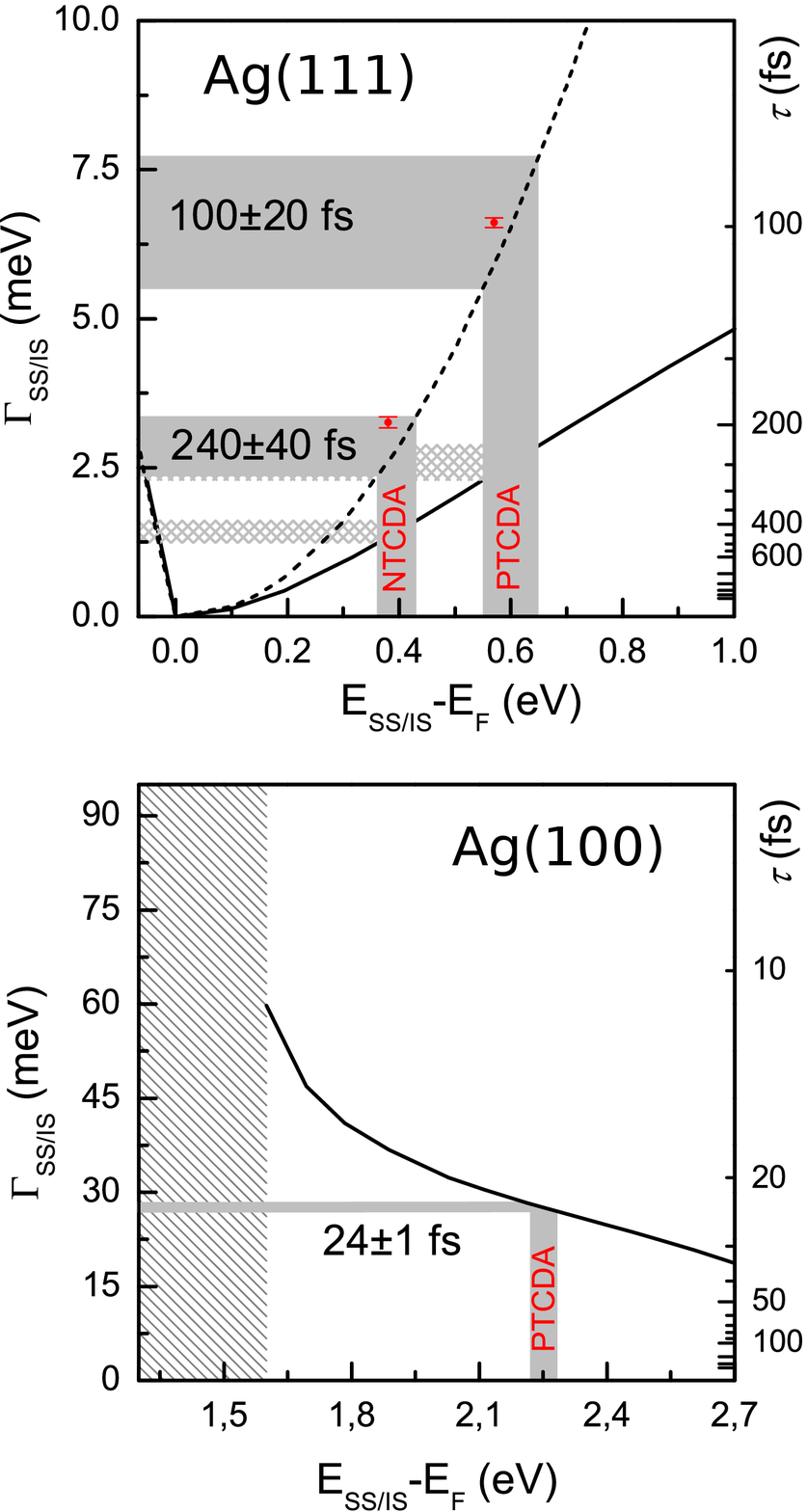}
\caption{Calculated decay rate $\Gamma$ of the shifted surface state  as a function of its energy. Solid (dashed) lines correspond to the $V$-($E$-)shifted scheme, respectively. Light-gray stripes cover the energy intervals, which correspond to the experimental values of the IS energy, including its variations with temperature. In the case of Ag(111), red points show the decay as obtained within the $E$-shifted scheme with taking into account the contribution of the former LUMO. Error bars here are caused by changing the former LUMO energy (see Fig.~\ref{fig:LCPAO}) within the $\pm50$ meV interval.\label{fig:Gamma}}
\end{center}
\end{figure}

In Fig.~\ref{fig:pseudo}, the starting-point 1D potential and the corresponding wave functions, as well as their counterparts at the largest energies considered, are shown. As seen in the figure, the shifted-SS penetration into the bulk becomes smaller upon modifying the potential, leading to a redistribution of the SS charge density into the vacuum side. 
This redistribution to a certain extent reflects the situation observed in \textit{ab initio} calculations, where in the case of the interface the charge outside the metal is larger than in the case of the bare surface (see Fig. \ref{fig:MLpot}).

On the one hand, the $V$-shifted scheme looks more consistent, because the energies $\epsilon_i$ and wavefunctions $\varphi_i(z)$ are the solutions of the same Schr\"odinger equation (\ref{schrodinger}). 
However, Fig.~\ref{fig:MLpot}(top) shows the ab initio calculated wave function of the IS in comparison with the SS of the bare silver surface, where along with the aforementioned redistribution to the vacuum side the penetration, and, consequently, overlapping with the bulk states, remains to be similar to the case of the bare surface. This means that the $E$-shifted SS scheme is valid and can be more suitable in the lifetime description.

As an advantage, both schemes developed allow us to get $\Gamma_{e-e}$ as a function of the IS energy, instead of specific calculations for a given IS energy.  
The decay rate of the  $V$-shifted SS electrons at $\bar{\Gamma}$ as a function of the SS(IS) energy is shown in Fig.~\ref{fig:Gamma}(top) for the Ag(111) surface. 
For example, for the experimental values of the IS energy observed at 300 K for NTCDA and PTCDA monolayers on Ag(111) (see Table~\ref{tab:IS_energies}) the decay rate $\Gamma_{e-e}=1.3$ meV ($\tau\sim$0.5 ps) and 2.4 meV ($\tau\sim$0.3 ps), respectively. Such a difference between the indicated values of $\Gamma_{e-e}$ is caused by the fact that for the bigger IS energy there is a larger decay phase space formed by the bulk states. However, the difference could be bigger, if the aforementioned overlap between the SS(IS) and the bulk states did not decrease with the increasing SS(IS) energy. Calculations performed with the unmodified bare-surface 1D potential but with changed SS(IS) energies ($E$-shifted surface state) clearly show how the decreasing overlap reduces the gain in the decay phase space (see dashed line in Fig.~\ref{fig:Gamma}(top)). In this case, for NTCDA and PTCDA monolayers on Ag(111) the decay rates are $\Gamma=2.6$ meV ($\tau\sim$0.3 ps) and 6.0 meV ($\tau\sim$0.1 ps), respectively. The absolute values of the lifetimes found within both schemes for the interfaces are substantially longer than those experimentally observed: 
$\tau^{IS}_{NTCDA@Ag(111)}=115\pm10$ fs and $\tau^{IS}_{PTCDA@Ag(111)}=53\pm3$ fs at 300 K and $\tau^{IS}_{NTCDA@Ag(111)}=43$ fs and $\tau^{IS}_{PTCDA@Ag(111)}=26$ fs at 90 K as reported in Ref.~\onlinecite{Marks_PRB_2011}. However, their ratio is very close to its experimental counterpart. Moreover, the presented results allow us to nicely reproduce the trend of the unusual increase of the lifetimes with rising temperature (see Fig.~\ref{fig:Gamma}(top)). Actually, at higher temperature the absorption distance becomes larger and, as a consequence, due to the weakened interaction between the molecular ML and metal substrate, the IS energy gets smaller, providing by this the smaller decay rate.

For the PFP@Ag(111) interface no experimental data is available on the IS lifetime. Our calculations yield a very small inelastic decay rate $\sim0.5$ meV, corresponding to the lifetime of 1-2 ps due to the low energy of the IS, and hence small phase space for the inelastic decay. However,  this value should be sufficiently shortened by elastic decay channels, e.g. electron-defect scattering.

In the case of the Ag(100) surface, the decay rate of the surface resonance (SR) cannot be calculated within the present framework. However, when the SR is pushed into the band gap, it becomes the IS. The decay rate of the latter  decreases with its energy (see Fig.~\ref{fig:Gamma}(bottom)) in spite of the growing decay phase space. Such a decrease is caused by the rapid reduction of the IS--bulk states overlap (see Fig.~\ref{fig:pseudo}). For the IS energy that corresponds to the PTCDA/Ag(100) interface, the decay rate is of 27 meV. The corresponding lifetime $\tau=24$ fs is surprisingly close to the upper limit of 18 fs found experimentally in Ref.~\onlinecite{Galbraith_JPCL_2014}.
Note that we consider here the $V$-shifted SS scheme only, since we cannot extrapolate the resonance-like wavefunction of the bare Ag(100) surface to the projected band gap energy region. 

\section{Monolayer pseudopotential \label{sec:MLpot}}
In an attempt to take into account a possible contribution coming from the molecular-derived states presented in Fig.~\ref{fig:LCPAO} by the fat bands (or red circles in Fig. \ref{fig:unfolded}a), the localization of these states lying energetically below the IS (the former LUMO in the Ag(111) interfaces) should be reproduced. (See Fig.~\ref{fig:MLpot}(top)) For this purpose we develop a pseudopotential that models the molecular ML as a quantum well and contains a barrier, separating the ML from the bulk Ag(111).  We refer to this model as {\it ML potential}.
This model allows us to get the shifted SS and the former LUMO with quite close energies (reproducing the ab initio values) and localized in the interface region and at the molecular layer, respectively, with the overlapping $\int|\varphi_{\rm IS}(z)|^2|\varphi_{\rm LUMO}(z)|^2dz$ as obtained from the \textit{ab initio} calculations.

In Fig.~\ref{fig:MLpot}(bottom), we plot the obtained model wave functions. Note, that in the calculation of lifetimes we have replaced the wavefunction of LUMO $\varphi_{\rm LUMO}(z)$ by $\varphi_{\rm LUMO}(z)/\sqrt{N}$ in order to take into account the unfolding  onto the ($1\times1$) Ag(111) BZ.
Here N is the number of atoms in the Ag(111) surface layer unit cell of the interface: $N=$ 33 and 24 for PTCDA and NTCDA, respectively.
 The resulting contribution accounting for transitions from the SS(IS) to the LUMO was estimated to be of $\sim0.3-0.7$ meV. Fig.~\ref{fig:Gamma}(top) shows the quite moderate effect of this contribution being added to the decay rate of the $E$-shifted SS scheme by red dots. By error bars, we demonstrate how variations in the former LUMO energy ($\pm50$ meV) can affect the resulting $\Gamma$. 

As one can see, the calculated lifetimes are generally longer, than the experimental ones. On the one hand, this is a typical picture, because there are usually some other decay channels, which are not taken into account in the model. On the other hand, the difference between theoretical and experimental results may be caused by inaccurate description of the charge density distribution. 
As follows from ab-initio calculations, in the interfaces under study the presence of the molecular ML causes a redistribution of not only the SS charge density in the vicinity of the ML plane, as we show above, but also the bulk-states charge density.
This simultaneous redistribution can produce an additional overlap between the IS and the bulk states, making the IS lifetime shorter. In order to take it into account, one should modify the 1D pseudo-potential in a way to properly describe its behavior within the ML region. Since it involves the bulk states and affects the image-potential tail of the 1D pseudo-potential, it should be done consistently with a study of energies and lifetimes of image-potential states, which, in turn, need to be experimentally analyzed first.
On the other side further experimental studies of IS lifetimes for the organic monolayers on top of the Ag(111) surface might shed light on peculiarities of this surface, that provide a distinct decay process as compared to the Ag(110) surface.

\section{Conclusions}

We have presented a theoretical study of lifetimes of interface states (IS) on metal-organic interfaces within one-dimensional pseudopotentials. 
The presented results allowed us to address the question about the description and explanation of recent experimental data.
We have demonstrated, that the folding of the BZ of the Ag surface due to sustantial enlarging of the surface unit cell upon the deposition of organic molecules does not have a drastic effect on the phase space of final states for electron decay. 

By means of the unfolding procedure we demonstrated, that the 
BZ folding upon organic monolayer deposition plays a minor role in the phase space for electron decay. Actually,  it introduces only a weak background to the bandstructure  of the ($1\times1$) unit cell, and hence weakly affects the lifetimes. 
The presence of the unoccupied molecular state below the IS gives a small contribution to the decay rate of the IS, while the lifetime  is mostly determined by the change of the phase space of final states upon the energy shift of the IS.

In the case of PTCDA@Ag(100), the  IS lifetime obtained in the $E$-shifted SS scheme agrees well with the experimental data, while for PTCDA and NTCDA at Ag(111) these model strongly overestimates the corresponding lifetimes of the IS. Being applied to these Ag(111) interfaces, the $E$-shifted SS scheme, which is based on the phase-space description, yields  shorter lifetimes, but still quite long as compared with the experiment. However, our calculations provide  the ratio $\tau^{IS}_{NTCDA@Ag(111)}/\tau^{IS}_{PTCDA@Ag(111)}$ in agreement with the experiments, and explain the trend of the unusual increase of of the IS lifetimes  with rising temperature.

\begin{acknowledgments} 

This work is a project of the SFB 1083 "Structure and Dynamics of Internal Interfaces" funded by the Deutsche Forschungsgemeinschaft (DFG).
We acknowledge partial support from the University of Basque Country UPV/EHU (IT-756-13),
the Departamento de Educaci\'on del Gobierno Vasco, The Tomsk State University Academic D.I. Mendeleev Fund Program (Grant No. 8.1.05.2015), the Spanish Ministry of Economy and Competitiveness MINECO (Grant No. FIS2013-48286-C2-1-P), and Saint
Petersburg State University (Project No. 11.50.202.2015).

\end{acknowledgments}



\begin{thebibliography}{00}

\bibitem{Szyman05pss} P. Szymanski, S. Garrett-Roe, and C.~B. Harris,
 Prog. Surf. Sci. {\bf 78}, 1 (2005).

\bibitem{Marks14jelsp} M. Marks, A. Sch{\"o}ll, and U. H{\"o}fer,
J. Electron. Spectrosc. Relat. Phenom. {\bf 195},  263  (2014).

\bibitem{Zhao14acsnano} J. Zhao, M. Feng, D.~B. Dougherty, H. Sun, and H. Petek,
 ACS Nano {\bf 8}, 10988  (2014).

\bibitem{Ilyas14prb} N. Ilyas and O. L. A. Monti,
  Phys. Rev. B {\bf 90}, 125435 (2014).

\bibitem{Temirov06nat} R. Temirov, S. Soubatch, A. Luican, and F.~S. Tautz,
 Nature {\bf 444},  350 (2006).

\bibitem{Schwalb08prl} C.~H. Schwalb, S. Sachs, M. Marks, A. Sch{\"o}ll, F. Reinert, E. Umbach,
    and U. H{\"o}fer, Phys. Rev. Lett. {\bf 101},  146801  (2008).

\bibitem{Yang08jpcc} A. Yang, S.~T. Shipman, S. Garrett-Roe, J. Johns, M. Strader, P. Szymanski,
    E. Muller, and C. Harris, J. Phys. Chem. C {\bf 112},  2506  (2008).

\bibitem{Scheybal09prb} A. Scheybal, K. Müller, R. Bertschinger, M. Wahl, A. Bendounan, P. Aebi,
    and T.~A. Jung, Phys. Rev. B {\bf 79},  115406  (2009).

\bibitem{Dyer_NJP_2010} M.~S. Dyer and M. Persson, New J. Phys. {\bf 12},  063014  (2010).

\bibitem{Zaitsev10} N.~L. Zaitsev, I.~A. Nechaev, and E.~V. Chulkov,
 J. Exp. Theor. Phys. {\bf 110},  114  (2010).

\bibitem{Sachs09jcp} S. Sachs, C.~H. Schwalb, M. Marks, A. Sch{\"o}ll, F. Reinert, E. Umbach, and
    U. H{\"o}fer, J. Chem. Phys. {\bf 131},  144701  (2009).

\bibitem{Schwalb10epj} C.~H. Schwalb, M. Marks, S. Sachs, A. Sch{\"o}ll, F. Reinert, E. Umbach,
    and U. H{\"o}fer, Eur. J. Phys. {\bf 75},  23  (2010).

\bibitem{Marks_PRB_2011} M. Marks, N.~L. Zaitsev, B. Schmidt, C.~H. Schwalb, A. Sch{\"o}ll, I.~A.
  Nechaev, P.~M. Echenique, E.~V. Chulkov, and U. H{\"o}fer,
  Phys. Rev. B {\bf 84},  081301(R)  (2011).

\bibitem{Zaitsev_PRB_2012} N.~L. Zaitsev, I.~A. Nechaev, P.~M. Echenique, and E.~V. Chulkov,
 Phys. Rev. B {\bf 85},  115301  (2012).

\bibitem{Galbraith_JPCL_2014} M.~C.~E. Galbraith, M. Marks, R. Tonner, and U. H{\"o}fer,
 J. Phys. Chem. Lett. {\bf 5},  50 (2014).


\bibitem{Caplins14jpcl} B.~W. Caplins, D.~E. Suich, A.~J. Shearer, and C.~B. Harris,
 J. Phys. Chem. Lett. {\bf 5},  1679  (2014).

\bibitem{Echenique04ssr} P.~M. Echenique, R. Berndt, E.~V. Chulkov, T.~H. Fauster, A. Goldmann, and
    U. H\"ofer, Surf. Sci. Rep. {\bf 52},  219  (2004).

\bibitem {hedin65} L.~Hedin, Phys. Rev. \textbf{139}, A796 (1965).
%
\bibitem{Chulkov_SS_1997} E.~V. Chulkov, V.~M. Silkin, P.~M. Echenique, Surf. Sci. {\bf391}, L1217 (1997).
%
\bibitem{Chulkov_SS_1999} E.~V. Chulkov, V.~M. Silkin, P.~M. Echenique, Surf. Sci. {\bf437}, 330 (1999).


%
\bibitem{_openmx_37} openmx-square.org.
%
\bibitem{ozaki_variationally_2003} T.~Ozaki, Phys. Rev. B \textbf{67}, 155108 (2003).
%
\bibitem{ozaki_numerical_2004} T.~Ozaki and H.~Kino, Phys. Rev. B \textbf{69}, 195113 (2004).
%
\bibitem{ozaki_efficient_2005} T.~Ozaki and H.~Kino, Phys. Rev. B \textbf{72}, 045121 (2005).
%
\bibitem{perdew_generalized_1996} J.~P. Perdew, K.~Burke, and M.~Ernzerhof, Phys. Rev. Lett. \textbf{77}, 3865 (1996).
%
\bibitem{troullier_efficient_1991} N.~Troullier, J.~L. Martins, Phys. Rev. B \textbf{43}, 1993 (1991).
%
\bibitem{stadler_molecular_2007} C.~Stadler, S.~Hansen, A.~Sch{\"o}ll, T.-L. Lee, J.~Zegenhagen, C.~Kumpf,
  and E.~Umbach, New Journal of Physics \textbf{9}, 50 (2007).
%
\bibitem{hauschild_normal-incidence_2010} A.~Hauschild, R.~Temirov, S.~Soubatch, O.~Bauer, A.~Sch{\"o}ll, B.~C.~C. Cowie, T.-L. Lee, F.~S. Tautz, and M.~Sokolowski, Phys. Rev. B \textbf{81}, 125432 (2010).
%
\bibitem{duhm_influence_2010} S.~Duhm, S.~Hosoumi, I.~Salzmann, A.~Gerlach, M.~Oehzelt, B.~Wedl, T.-L. Lee,
  F.~Schreiber, N.~Koch, N.~Ueno, and S.~Kera, Phys. Rev. B \textbf{81}, 045418 (2010).


\bibitem{unfolding1} BandUP: Band unfolding code for plane-wave based calculations, \href{www.ifm.liu.se/theomod/compphys/band-unfolding}{www.ifm.liu.se/theomod/compphys/band-unfolding}.

\bibitem{unfolding2}
P.V.C.~Medeiros, S.~Stafstr\"om and J.~Bj\"ork, Phys. Rev. B {\bf89}, 041407(R) (2014).

\bibitem{unfolding3}
P.V.C.~Medeiros, S.S.~Tsirkin, S.~Stafstr\"om and J.~Bj\"ork, Phys. Rev. B. {\bf91}, 041116(R) (2015).

\bibitem{Popescu}
V.~Popescu and A.~Zunger, Phys. Rev. B {\bf85}, 085201 (2012).

\bibitem{VASP1} G.~Kresse and J.~Hafner, Phys. Rev. B {\bf 48}, 13115 (1993).
\bibitem{VASP2} G.~Kresse and J.~Furthm\"{u}ller, Comput. Mater. Sci. {\bf 6}, 15 (1996).
\bibitem{PAW1} P.E.~Bl\"{o}chl, Phys. Rev. B {\bf 50}, 17953 (1994).
\bibitem{PAW2} G.~Kresse and D.~Joubert, Phys. Rev. B {\bf 59}, 1758 (1999).

%
\bibitem{Galbraith_PHD} M.C.E. Galbraith, M.Sc. thesis, Philipps-Universit\"{a}t Marburg, 2012.
%
\bibitem{Kliewer_Science_2000} J. Kliewer, R. Berndt, E. V. Chulkov, V. M. Silkin, P. M. Echenique, S. Crampin, Science {\bf288}, 1399 (2000).
%
\bibitem{Vitali_SSL_2003} L. Vitali, P. Wahl, M. A. Schnider, K. Kern, V. M. Silkin, E. V. Chulkov, P. M. Echenique, Surf. Sci. Lett. {\bf523}, L47 (2003).


\end{thebibliography}
\end{document}